\documentclass{publisher/svproc}

\usepackage{amstext}               
\usepackage[greek,english]{babel}  
\usepackage[utf8x]{inputenc}       

\newcommand{\MACH}[1]{\text{\scshape\mdseries\rmfamily #1}}
\newcommand{\MACHNUM}[1]{\oldstylenums{\upshape#1}}
\newcommand{\MACHs}{\text{\upshape\mdseries\rmfamily s}}

\newcommand{\fa  }{\MACH{fa}}  

\newcommand{\MACHdeterministic         }{\MACH{d}}

\newcommand{\MACHnondeterministic      }{\MACH{n}}

\newcommand{\MACHoneway   }{\MACHNUM{1}}

\newcommand{\MACHtwoway   }{\MACHNUM{2}}

\newcommand{\onfa   }{{\MACHoneway\MACHnondeterministic\fa}}

\newcommand{\onfas  }{{\onfa\MACHs}}

\newcommand{\tdfa   }{{\MACHtwoway\MACHdeterministic\fa}}

\newcommand{\tnfa   }{{\MACHtwoway\MACHnondeterministic\fa}}

\newcommand{\tdfas  }{{\tdfa\MACHs}}

\newcommand{\tnfas  }{{\tnfa\MACHs}}

\usepackage{amssymb}        
\usepackage{amsmath}        

\newcommand{\bbB}{\mathbb{B}}

\newcommand{\ggS}{\varSigma}

\newcommand{\ggV}{\varOmega}

\newcommand{\gga}{\alpha}
\newcommand{\ggb}{\beta}

\newcommand{\ggi}{\iota}

\newcommand{\kkk}{{,}}

\newcommand{\amp}{\;\,\textup{\&}\;\,}

\newcommand{\ppp}{{+}}
\newcommand{\mmm}{{-}}
\newcommand{\xxx}{{\times}}

\newcommand{\atmost}{{\leq}\,}
\newcommand{\atleast}{{\geq}\,}

\newcommand{\subs }{\subseteq }
\newcommand{\sups }{\supseteq }
\newcommand{\ssubs}{\subsetneq}

\newcommand{\nsubs}{\nsubseteq}
\newcommand{\nsups}{\nsupseteq}
\newcommand{\unicup}{{\textstyle\bigcup}}

\newcommand{\ppows}[1]{\mbox{$[\hspace{-1.5pt}[$}#1\mbox{$]\hspace{-1.5pt}]$}}

\newcommand{\poly}{\text{\upshape\mdseries poly}}

\newcommand{\s}{^*}
\newcommand{\z}{_*}

\newcommand{\fml}[2][1]{(#2_h)_{h\geq#1}}

\newcommand{\tagLR  }{\textup{\textsc{lr}}}
\newcommand{\tagRL  }{\textup{\textsc{rl}}}

\usepackage{graphicx}             
\usepackage[usenames]{color}      

\newcommand{\figdir}{figures}

\graphicspath{{\figdir/}}

\usepackage{amstext}        
\usepackage{xstring}        

\newcommand{\PROB    }[1]{\text{\scshape\mdseries\rmfamily #1}}
\newcommand{\PROBMATH}[1]{\text{\scriptsize$#1$}}

\newcommand{\nowl     }[1][n]{\IfInteger{#1}{\PROBMATH{#1}}{\ensuremath{#1}}\PROB{owl}}

\newcommand{\owl            }{\PROB{owl}}

\usepackage{amstext}               
\usepackage[greek,english]{babel}  
\usepackage[utf8x]{inputenc}       

\newcommand{\CLSS   }[1]{\text{\upshape\mdseries\sffamily #1}}
\newcommand{\CLSSNUM}[1]{\CLSS{#1}}

\newcommand{\CLSSnondeterministic      }{\CLSS{N}}

\newcommand{\CLSSoneway   }{\CLSSNUM{1}}

\newcommand{\onen   }{{\CLSSoneway\CLSSnondeterministic}}

\newcommand{\p      }{\CLSS{P}}                                 
\newcommand{\np     }{\CLSS{NP}}                                

\renewcommand{\qed}{\hfill$\Box$}
\newenvironment{myproof}
  {\begin{proof}}
  {\qed\end{proof}}
\newenvironment{myproofnoqed}
  {\begin{proof}}
  {\end{proof}}

\newcommand{\qqed}{\hfill$\boxdot$}
\newenvironment{mypproof}
  {\begin{proof}}
  {\qqed\end{proof}}
\newenvironment{mypproofnoqed}
  {\begin{proof}}
  {\end{proof}}

\usepackage{endnotes}

\newcommand{\notesskip}{\bigskip}

\renewcommand{\notesname}{Notes}
\newcommand{\notesnameforsubmission}{
 \renewcommand{\notesname}{Notes\hfill{\scriptsize(\textit{To be omitted in final version.})}}}

\newcommand{\myendnote}[1]{\endnote{#1\notesskip}}
\newcommand{\myendnotetext}[2][0]{%
    \addtocounter{endnote}{#1}%
    \endnotetext{#2\notesskip}}
\newcommand{\myendnoteproof}[3]{\endnotetext{%
    \textsl{Proof of #1~#2.}\par#3\qed\notesskip}}

\newcommand{\volume}{\mathrm{vol}}
\newcommand{\closure}{\mathrm{cl}}
\newcommand{\height}{\mathrm{h}}
\newcommand{\core}[1]{\mathrm{core}(#1)}

\begin{document}

\mainmatter
\title{Unrestricted 2DFA simulation of 1NFAs: A~Quadratic Limitation to a New Lower Bound}
\titlerunning{A limitation for proving lower bounds for 2DFAs against OWL}
\author{Kehinde Adeogun \and Christos A.\ Kapoutsis}
\authorrunning{K.~Adeogun \and C.\,A.~Kapoutsis} 
\institute{Carnegie Mellon University in Qatar}
\maketitle

\begin{abstract}
\!A recent result by the present authors established a quadratic lower bound, in the worst case, for the increase in the number of states when a \textit{one-way nondeterministic finite automaton} is converted to a \textit{two-way deterministic finite automaton}. Although this simply matched a well-known pre-existing quadratic lower bound by Chrobak, it used a distinct proof method. We show that, much like Chrobak's, this new method is also unable to deliver any lower bound strictly greater than quadratic. 
\end{abstract}

\section{Introduction}
\label{sec:introduction}

An important open question in automata and complexity theory concerns the conversion of \textit{one-way nondeterministic finite automata} (\onfas) to \textit{two-way deterministic finite automata} (\tdfas). By an old result~\cite{rasc59}, this conversion is always possible, as we can convert the given \onfa~$N$ to an equivalent \textit{one-way} deterministic finite automaton~$M$, i.e., a \tdfa\ that does not use its ability to reverse its input head. However, this method is known to incur an exponential cost in the number of states: if $N$~has $s$~states, then $M$~may well have (and, for some~$N$, necessarily has) $2^s$~states~\cite{mefi71}. A natural question then is whether $M$~can decrease its number of states to only a polynomial in~$s$, if it uses its bidirectionality.  

First posed by Seiferas~\cite{se73}, this question is still open. Sakoda and Sipser~\cite{sasi78} revealed its importance by viewing it as a special case of the respective question for the more general conversion of \textit{two-way} nondeterministic finite automata (\tnfas) to \tdfas; and, in turn, by viewing this latter conversion as a miniature variant of our big open questions on the power of nondeterminism (e.g., \p~vs.~\np). In the spirit of this analogy, they also introduced a decision problem, now called \textit{one-way liveness} (\owl), that is ``complete'' for the conversion, in that: for all~$h$, a \tdfa\ with $s(h)$~states can decide $\owl_h$ (i.e., \owl\ on instances of ``height''~$h$) iff every $h$-state \onfa\ has an equivalent \tdfa\ with $s(h)$~states. Hence, the original question became: Is $\owl_h$ decidable by a \tdfa\  with $\poly(h)$~states?

Sakoda and Sipser conjectured a negative answer, namely that \tdfas\ need super-polynomially many states to decide $\owl_h$; and thus (equivalently) to simulate $h$-state \onfas\ (and thus to simulate $h$-state \tnfas---this latter, weaker claim being what is now known as the ``Sakoda-Sipser conjecture''). Their first attempts at confirming this conjecture were followed by many others, which mostly established super-polynomial lower bounds for special cases of \textit{restricted} \tdfas\ (see~\cite{adka26sofsem}\nocite{adka26dmtcs} for a brief summary). Proving lower bounds for \textit{unrestricted} \tdfas\ against \owl\ has been a minority approach, with just three instantiations.

In~1986, Chrobak~\cite{ch86}\nocite{to09} studied the conversion in the {special case} where the given \onfa\ is \textit{unary} (i.e., over a single-letter input alphabet). For every~$h$, he exhibited a specific $h$-state \onfa~$N_h$ and proved that every \tdfa\ equivalent to it needs $\ggV(h^2)$~states, directly implying the same for the restriction of~$\owl_h$ to strings of the one symbol that captures the behavior of that~$N_h$. On a pessimistic note, he also proved that every $h$-state unary \onfa\ has an equivalent \tdfa\ with $O(h^2)$~states, so his approach was unable to deliver any greater lower bound.  

In~2018, Kapoutsis~\cite{ka18}\nocite{bokoun18} studied $\owl_h$ in the {special case} of instances of length~$3$. He proved that every \tdfa\ deciding such instances needs $\ggV(h^2/\log h)$ states. On a pessimistic note, he also proved that $O(h^2/\log h)$ states are enough, so his approach was also unable to deliver any greater lower bound.

Note that, quite counter-intuitively, in both of the above results the lower bound was achieved even on severely restricted instances of~\owl\ (strings over only one symbol; or of length~$3$). However, as mentioned above, the \tdfa\ assumed to be working on these restricted instances was completely unrestricted.

In~2026, the present authors~\cite{adka26sofsem} studied $\owl_h$ in the {general case}. Matching Chrobak's lower bound, they proved that every \tdfa\ for~$\owl_h$ must have at least~$\frac12\binom{h+1}{2}=\ggV(h^2)$ states. Crucially, their technique was distinct from those of~\cite{ch86,ka18}: they proved that, under certain favorable conditions, two  \textit{properties}~$P,P'$ (i.e., sets of strings) can force the \tdfa\ to ``spend $1$~state on the left or right'' just to be able to tell whether some infix of the input is in~$P$ or~$P'$; and then designed a sequence of properties $P_0,P_1,\dots,P_N$ where $N=\binom{h+1}{2}$ and every $P_{i-1},P_i$ satisfy the favorable conditions, thus concluding that $\frac12N$~states must be spent. On a pessimistic note, they also \textit{conjectured} that their sequence $P_0,P_1,\dots,P_N$ was longest among all sequences where successive properties satisfy the favorable conditions; so their approach was also unable to deliver any greater lower bound. 

Here, we confirm that conjecture (cf.~Th.~\ref{thm:main}). We prove that, indeed, for every sequence $P_0,P_1,\dots,P_m$, if every two successive properties $P_{i-1},P_i$ satisfy the favorable conditions of~\cite[Lemma~10]{adka26sofsem} (for $\owl_h$), then $m\leq N=\binom{h+1}{2}$. For this, we first abstract every instance~$z$ of $\owl_h$ into its \textit{connectivity}~$C$: an $h\times h$ Boolean matrix that records connections between nodes of the outer columns of~$z$. We then view~$C$ as a \textit{homomorphism} on the powerset of~$[h]:=\{1,\dots,h\}$, and note that its range is a \textit{lattoid}~$V$: a union-closed family of subsets of~$[h]$. In it, we identify a quantity, the \textit{prime volume} of~$V$, which lies between $0$ and~$\binom{h+1}{2}$. Finally, we show that, in every step along the given sequence $P_0,P_1,\dots,P_m$, the prime volume decreases by at least~$1$, which immediately implies $m\leq\binom{h+1}{2}$.

After some preliminaries (Sect.~\ref{sec:preliminaries}), we study lattoids and how mappings on them affect the prime volume (Sect.~\ref{sec:lattoids}); connectivities and how they relate to lattoids (Sect.~\ref{sec:connectivities}); and properties and how they relate to connectivities (Sect.~\ref{sec:connectivity-properties}). We then prove our result for the special case of \textit{connectivity properties} (Th.~\ref{thm:main-conn}), like those designed in~\cite{adka26sofsem}. The general case (Th.~\ref{thm:main}) needs a simple extension with longer presentation and is reserved for the full version of this report.

\newpage 

\section{Preliminaries}
\label{sec:preliminaries}

For $h\geq1$, let $[h]:=\{1,2,\ldots,h\}$ and let~$\ppows{h}$ be the powerset of~$[h]$. In a family of sets $F\subs\ppows h$, an \textit{incomparability} is any unordered pair $\{a\kkk b\}$ of \textit{incomparable} sets: $a,b\in F \amp a\nsubs b \amp a\nsups b$. The set of all incomparabilities in~$F$ is denoted by~$I(F)$. The \textit{volume} of~$F$ is its number of sets and incomparabilities
\begin{equation}
\volume(F):=|F|+|I(F)|\,,
\end{equation}
and is between~$0$ and $\binom{|F|+1}{2}$.%
\myendnote{\label{end:bounds-for-volume}%
The lower bound is trivial, for a sum of non-negative terms; and is clearly reached when $F=\emptyset$. For the upper bound, the second term is clearly at most the number $\binom{|F|}{2}$ of all unordered pairs in~$F$, so the sum is at most $|F|+\smash{\binom{|F|}{2}}$, which evaluates to $\smash{\binom{|F|+1}{2}}$; moreover, this bound is clearly met when every two sets are incomparable, namely when $F$ is an anti-chain.}
The \textit{union closure} of~$F$ is the set of all unions of (zero or more) sets of~$F$, $\closure(F):=\{\unicup X\mid X\subs F\}$. Easily, this includes the empty set~$\emptyset$;%
\myendnote{\label{end:union-closure-contains-emptyset}%
Indeed, choosing from~$F$ the empty collection of sets $X:=\emptyset\subs F$, we have that $\unicup X\in F$, where $\unicup X=\unicup\emptyset=\emptyset$. Hence, $\emptyset\in F$.}
and all of~$F$,%
\myendnote{\label{end:union-colsure-contains-all}%
Indeed, for every $a\in F$, choosing from~$F$ the singleton collection $X:=\{a\}\subs F$, we have that $\unicup X\in F$, where $\unicup X=\unicup\{a\}=a$. Hence, $a\in\closure(F)$.} 
i.e., $F\subs\closure(F)$. If $F=\closure(F)$, then $F$~is \textit{union-closed}. 

\begin{example}
Family $F:=\bigl\{\{1\},\{2\},\{2\kkk3\}\bigr\}$ has two incomparabilities ($\{1\}$ with $\{2\}$ and~$\{2\kkk3\}$); volume $3\ppp2=5$; and union closure~$\ppows3\setminus\bigl\{\{3\},\{1\kkk3\}\bigr\}$.
\end{example}

\paragraph{Properties.}
\label{sec:properties}

A \textit{property} (of the strings) over an alphabet~$\ggS$ is any $P\subs\ggS\s$. We call~$P$ \textit{smooth}, if every two strings in it have an infix that keeps their concatenation in the property: $(\forall x\kkk z\in P)(\exists y)(xyz\in P)$. Properties $P,P'$ are \textit{separated} by language~$L\subs\ggS\s$ if there is always a context where choosing $P$ or $P'$ affects membership in~$L$: $(\forall x\in P)(\forall z\in P')(\exists u\kkk v)(uxv\in L \iff uzv\not\in L)$.

We say that $P$~\textit{\tagLR-extends} to~$P'$ if every string in it can be \tagLR-extended into~$P'$: $(\forall x\in P)(\exists y)(xy\in P')$; that \textit{$P$~has a suffix into~$P'$}, if a single suffix~$y$ can push into~$P'$ every~$x$ from~$P$ or~$P'$: $(\exists y)(\forall x\in P\cup P')(xy\in P')$; and that \textit{$P$~has suffix of choice into~$P'$}, if that suffix can be chosen to end in any way $v\in P$ that we want: $(\forall v\in P)(\exists u)(\forall x\in P\cup P')(xuv\in P')$.

\begin{figure}[t]
\centering
\begin{tabular}{ccc}
\raisebox{0.00cm}{\includegraphics[height=2.4cm]{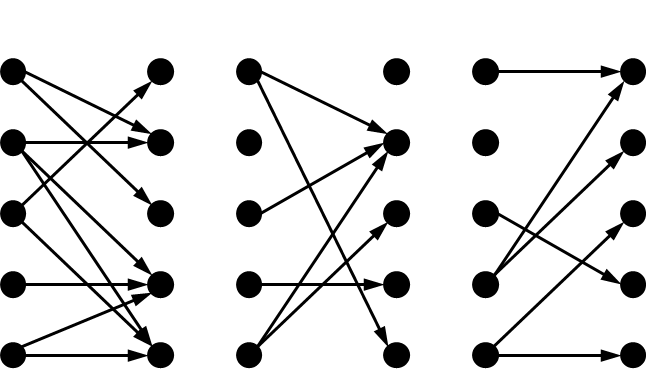}} & \qquad\quad
\raisebox{0.00cm}{\includegraphics[height=2.4cm]{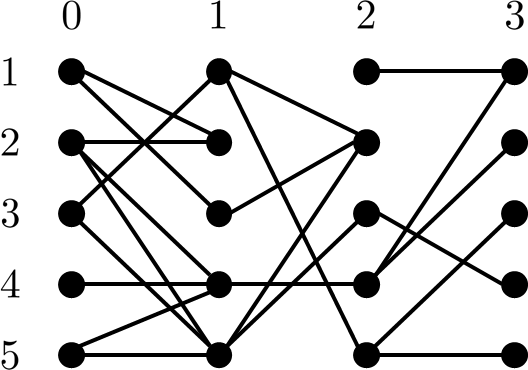}} & \qquad\quad
\raisebox{0.9cm}{
$\begin{bmatrix}
0 & 0 & 0 & 0 & 0\\
1 & 1 & 0 & 1 & 0\\
0 & 0 & 1 & 1 & 1\\
1 & 1 & 0 & 0 & 0\\ 
1 & 1 & 0 & 1 & 0\\ 
\end{bmatrix}$}
\end{tabular}
\caption{Three symbols in~$\ggS_5$~({\textsf{\scriptsize L}}); their string~({\textsf{\scriptsize M}}); and its connectivity~({\textsf{\scriptsize R}}). (From~\cite{adka26sofsem}.)}
\label{fig:owl}
\end{figure}

\paragraph{One-way liveness.}
\label{sec:owl}

Let $h\geq1$. The alphabet $\ggS_h:=\mathcal{P}([h]\xxx[h])$ consists of every two-column directed graph with $h$~nodes per column and only rightward arrows (Fig.\,\ref{fig:owl}). An $n$-long string $z\in\ggS\s_h$ is naturally viewed as a graph of $n\ppp1$~columns, indexed from~$0$ to~$n$, where edges connect successive columns and, for simplicity, are undirected. If there exists an $n$-long path from some node of column~$0$ to some node of column~$n$, then $z$~is \textit{live}; otherwise, it is \textit{dead}. Checking that a given $z\in\ggS_h\s$ is live is the \textit{one-way liveness} problem for height~$h$, denoted by~$\owl_h$. 

The problem helps us study how easily \tdfas\ can simulate \onfas~\cite{sasi78}. Formally, the family $\owl=\fml\owl$ is complete, under homomorphic reductions, for the class \onen\ of all language families $L=\fml L$ that are recognized by \onfa\ families $N=\fml N$ where the number of states grows polynomially in~$h$. Intuitively, the strings of $\ggS_h\s$ are all computations of all $h$-state \onfas. So, if a \tdfa\ solves $\owl_h$ with $\atmost s(h)$~states, then every $h$-state \onfa\ is simulated by a \tdfa\ with $\atmost s(h)$ states; and, conversely, if every \tdfa\ for $\owl_h$ needs $\atleast s(h)$~states, then some $h$-state \onfa\ needs $\atleast s(h)$ states in each \tdfa~simulator.

\paragraph{The new method.}
\label{sec:the-new-method}

At the heart of the method of~\cite{adka26sofsem} lies a general lemma, that a \tdfa\ for a language~$L$ must ``spend $\atleast 1$ state on the left or the right'' just to ``distinguish'' between two properties $P$ and~$P'$, if these satisfy certain conditions. 

\begin{lemma}[\textmd{\cite[Lemma\,10]{adka26sofsem}}]\label{lem:AK26}
Let $P,P'\neq\emptyset$ be smooth. Suppose $P$~has suffix of choice into~$P'$\textup; and $P,P'$ are separated by~$L$. If \tdfa~$M$ solves~$L$, then the exit sizes of $P,P'$ on $M$ satisfy $a(P)\geq a(P')$ and\/ $b(P)\geq b(P')$, and at least one of these inequalities is strict.
\end{lemma}

\noindent
For the definitions of  \tagLR-\textit{size}~$a(P)$ and \tagRL-\textit{size} $b(P)$ of a property~$P$, see~\cite{adka26sofsem}. Here, we only need to know that all of the mentioned \tagLR-/\tagRL-sizes are between~$0$ and the size of $M$'s set of states~$Q$, namely: $0\leq a(P),b(P),a(P'),b(P')\leq|Q|$.

Given this lemma for $L=\owl_h$, the rest of~\cite{adka26sofsem} is spent designing properties $P_0,P_1,\dots,P_N\subs\ggS_h\s$ for $N=\binom{h+1}{2}$ such that every two successive ones $P_{i-1},P_i$ satisfy the lemma's assumptions (non-empty, smooth, separated by~$\owl_h$, with suffix of choice) and thus its conclusion  that $a(P_{i-1})\geq a(P_i) \amp b(P_{i-1})\geq b(P_i)$ and at least one inequality is strict. Therefore, the two sequences
\begin{equation}
|Q|\geq a(P_0) \geq \cdots \geq a(P_N) \geq 0
\quad\text{and}\quad
|Q|\geq b(P_0) \geq \cdots \geq b(P_N) \geq 0
\end{equation}
host $\atleast N$ strict inequalities; so one of them hosts $\atleast \frac12N$; therefore $|Q|\geq \frac12N$. 

Our goal is to prove that the sequence $P_0,P_1,\dots,P_N$ above is longest, and so this method can deliver no lower bound strictly greater than $\frac12N=\frac12\binom{h+1}{2}$. 

\begin{theorem}\label{thm:main}
Let $\emptyset\neq P_0,P_1,\dots,P_m\subs\ggS_h\s$ be arbitrary properties. Suppose that, for~$1\leq i\leq m$, 
$P_{i-1},P_i$~are separated by~$\owl_h$\textup; $P_{i-1}$~has suffix of choice into~$P_i$\textup; and $P_i$~is smooth. Then $m\leq\binom{h+1}{2}$.
\end{theorem}

\noindent 
As already mentioned, our proof here will be only for when all~$P_i$ are \textit{connectivity properties} (cf.\ Sect.~\ref{sec:connectivity-properties}); the full case will be proved in the full version of our report.

\section{Lattoids}
\label{sec:lattoids}

A \textit{lattoid} over~$h$ is any union-closed family $V\subs\ppows h$. Note that such~$V$ always contains~$\emptyset$.%
\myendnote{\label{end:lattoid-contains-emptyset}%
It is $V=\closure(V)$ (as $V$ is union-closed) and $\emptyset\in\closure(V)$ (Endnote~\ref{end:union-closure-contains-emptyset}); so, $\emptyset\in V$.}
For~$v\in V$, the length of a longest strict chain in~$V$ from~$\emptyset$ to~$v$:
\begin{equation}
\height(v):=\max\{ k \mid (\exists v_0,v_1,\dots,v_k\in V)(\emptyset=v_0\ssubs v_1\ssubs\cdots\ssubs v_k =v)\} \,,
\end{equation}
 is its \textit{height}. Easily, we have $0=\height(\emptyset)\leq \height(v)$ and $u\ssubs v\implies \height(u)<\height(v)$.%
\myendnote{\label{end:height-is-strictly-increasing}%
Suppose $u\ssubs v$. Let $k:=\height(u)$ and consider any strict chain that witnesses this: $\emptyset=u_0\ssubs u_1\ssubs\cdots\ssubs u_k=u$. Then the chain $u_0\ssubs u_1\ssubs\cdots\ssubs u_k=u\ssubs v$ proves that $k\ppp1\leq \height(v)$. So, $\height(u)<\height(v)$.}

We say that $v$~is \textit{composite}, if it is the union of \emph{other} sets in $V$, namely if 
$(\exists X \subseteq V)\bigl( v\notin X \amp v=\unicup X \bigr)$;
otherwise, it is~\emph{prime}. Note that $\emptyset$~is composite.%
\myendnote{Indeed, if $X=\emptyset\subs V$, then we have $\emptyset=\unicup X$ and $\emptyset\notin X$.}
The set of all primes in~$V$ is its \textit{core}, $\core V$. A \textit{prime factorization} of~$v$ is any collection~$X$ of primes that can create it: $X\subs\core V \amp v=\unicup X$. Easily, prime factorizations always exist.
\begin{lemma}\label{lem:prime-factorization-exists}
Every set~$v$ in a lattoid~$V$ has a prime factorization.\endnotemark
\end{lemma}
\myendnoteproof{Lemma}{\ref{lem:prime-factorization-exists}}{%
By strong induction on $k:=\height(v)$. In the base case $(k=0)$, it can only be $v=\emptyset$, which has prime factorization $X_\emptyset:=\emptyset$, since $X_\emptyset\subs\core V$ and $\unicup X_\emptyset=\emptyset$. 
\par
For the inductive set $(k\geq1)$, consider any $v\in V$ with height~$k$. \textit{If $v$~is prime}, then $X_v:=\{v\}$ is clearly a prime factorization. \textit{Otherwise}, $v$~is composite, namely the union of a collection~$X\subs V$ of  \textit{other} sets: $v\notin X \amp v=\unicup X$. Then each $u\in X$ has strictly smaller height than~$v$ (see below) and thus (by inductive hypothesis) a prime factorization $X_u\subs\core V$ with $u=\unicup X_u$. Collect all primes from all these factorizations into $X_v:=\bigcup_{u\in X}X_u$. Then easily
\begin{equation}
\bigcup X_v = 
\bigcup \bigl( \bigcup_{u\in X} X_u \bigr) = 
\bigcup_{u\in X} \bigl( \bigcup X_u \bigr)= 
\bigcup_{u\in X} u = 
\bigcup X = 
v;
\end{equation}
and clearly $X_v\subs\core V$; hence, $X_v$ is a prime factorization of~$v$, as desired.
\par
To see why each $u\in X$ has strictly smaller height than~$v$, note that $u\subs v$ (since $u\subs\unicup X=v$) and $u\neq v$ (since $v\notin X$). Therefore $u\ssubs v$, which implies $\height(u)<\height(v)$ (Endnote~\ref{end:height-is-strictly-increasing}).}

The \textit{prime incomparabilities} and \textit{prime volume} of a lattoid~$V$ are the incomparabilities and volume of its core. Easily, $0\leq\volume(\core V)\leq\binom{|\core V|+1}{2}$.\label{loc:bounds-for-prime-volume}\endnotemark\
\myendnotetext{\label{end:bounds-for-prime-volume}%
Follows from the bounds $0\leq\volume(F)\leq\binom{|F|+1}{2}$ for arbitrary~$F\subs\ppows h$ (Endnote~\ref{end:bounds-for-volume}), when we let $F:=\core V$. Easily, the lower bound is met when $\core V=\emptyset$, equivalently when $V$ is the singleton~$\{\emptyset\}$; and the upper bound is met when the primes of~$V$ form an anti-chain.}

A \textit{representative} for a prime~$v$ of a lattoid~$V$ is any member~$i$ that belongs to only those primes of~$V$ that contain~$v$: 
\begin{equation}\label{equ:representative}
i\in v 
\quad\amp\quad
(\forall u\in\core V)( i\in u \implies v\subs u) \,.
\end{equation}
We also say that $i$~\textit{represents}~$v$; and $v$~is \textit{represented} (by~$i$). Easily, a prime may have zero, one, or many representatives; but each~$i$ can represent at most one prime in~$V$.\label{loc:each-rep-at-most-one-prime}%
\myendnote{\label{end:each-rep-at-most-one-prime}%
Suppose $i$~represents primes $v$ and~$v'$. Then \eqref{equ:representative}~implies $i\in v$ and $i\in v'$. Also, because $i$~represents~$v$ and is in~$v'$, \eqref{equ:representative}~implies $v\subs v'$. Likewise, because $i$~represents~$v'$ and is in~$v$, \eqref{equ:representative}~implies $v'\subs v$. So, $v=v'$.}
If every prime is represented, we say that $V$~is a \textit{regular} lattoid.

\subsection{Mappings between Lattoids}
\label{sec:mappings}

Let~$U,V$ be lattoids and $\gga: U\to V$ any mapping. We say that $\gga$~\textit{respects inclusion} (or is \textit{monotone}) if $u\subs u'\implies \gga(u)\subs\gga(u')$; that $\gga$~\textit{respects non-inclusion} if $u\nsubs u'\implies \gga(u)\nsubs \gga(u')$; that $\gga$~is an \textit{embedding}, if it respects both inclusion and non-inclusion; and that $\gga$~is a (\textit{relational}) \textit{isomorphism} if it is a bijective embedding. Easily, surjective embeddings are already isomorphisms: 

\begin{lemma}\label{lem:surj-embed-is-rel-iso}
Every surjective embedding is a \textup(relational\textup) isomorphism.\endnotemark
\end{lemma}
\myendnoteproof{Lemma}{\ref{lem:surj-embed-is-rel-iso}}{%
Let $\gga:U\to V$ be a surjective embedding. We just prove that $\gga$~is injective (as then it is a bijective embedding, i.e., a relational isomorphism). Suppose $u\neq u'$. Then $u\nsubs u'$ or $u\nsups u'$. Without loss of generality, assume $u\nsubs u'$. Then $\gga(u)\nsubs\gga(u')$ (since $\gga$ respects non-inclusion). So, $\gga(u)\neq\gga(u')$.}

We say that $\gga$~is a \textit{homomorphism} (or that it \textit{respects union}) if, for all $X\subs U$, it is $\gga(\unicup X)=\bigcup_{u\in X}\gga(u)$. Easily, every homomorphism is a monotone mapping; is fully determined by its values on the primes; and its range is a lattoid:

\begin{lemma}\label{lem:homoms-are-monotone}
Every homomorphism respects inclusion.\endnotemark
\end{lemma}
\myendnoteproof{Lemma}{\ref{lem:homoms-are-monotone}}{%
Let $\gga:U\to V$ be a homomorphism and $u,u'\in U$, Suppose $u\subs u'$. Then clearly $u\cup u'=u'$; and thus $\gga(u')=\gga(u\cup u')=\gga(u)\cup\gga(u')$ (since $\gga$~is a homomorphism); which clearly contains $\gga(u)$. Overall, $\gga(u)\subs\gga(u')$.}

\begin{lemma}\label{lem:values-on-core-determine-homomorphism}
Suppose homomorphisms $\gga,\ggb:U\to V$ agree on the primes of~$U$, namely $\gga(u)=\ggb(u)$ for every~$u\in\core U$. Then $\gga=\ggb$.\endnotemark
\end{lemma}
\myendnoteproof{Lemma}{\ref{lem:values-on-core-determine-homomorphism}}{%
Let $v\in U$. Let $X_v\subs\core U$ be a prime factorization of~$v$ (Lemma~\ref{lem:prime-factorization-exists}). Then  $v=\unicup X_v$ and we calculate:
\begin{equation}
\gga(v) = 
\gga(\unicup X_v) \stackrel{(1)}{=} 
\bigcup_{u\in X_v} \gga(u) \stackrel{(2)}{=}
\bigcup_{u\in X_v} \ggb(u) \stackrel{(3)}{=}
\ggb(\unicup X_v) =
\ggb(v) \,,
\end{equation}
where (1) and (3)~use the fact that $\gga,\ggb$ are homomorphisms; and (2)~uses their agreement on the primes.}

\begin{lemma}\label{lem:range-of-homom-is-lattoid}
The range of every homomorphism is a lattoid.\endnotemark
\end{lemma}
\myendnoteproof{Lemma}{\ref{lem:range-of-homom-is-lattoid}}{%
Let $\gga:U\to V$ be a homomorphism. Let $\tilde U:=\{\gga(u)\mid u\in U\}$ be its range. To prove that $\tilde U$ is a lattoid, we pick an arbitrary collection of sets $\tilde X\subs\tilde U$ and show that their union $\tilde u:=\unicup\tilde X$ is also in~$\tilde U$, as follows. 

Let $X:=\{ u\in U \mid \gga(u)\in\tilde X\}$ consist of all sets in~$U$ that map to sets of~$\tilde X$. Then their images are exactly $\tilde X$, namely $\{\gga(u)\mid u\in X\} = \tilde X$. Therefore:
\begin{equation}\textstyle
\tilde u =
\unicup\tilde X = 
\unicup\{\gga(u)\mid u\in X\} =
\bigcup_{u\in X} \gga(u)  \stackrel{(1)}{=}
\gga( \bigcup_{u\in X} u ) =
\gga(\unicup X) \,,
\end{equation}
where (1)~uses the fact that $\gga$~is homomorphism. Hence, $\tilde u$ is the $\gga$-image of the set $\unicup X\in U$, which implies $\tilde u\in\tilde U$, as required.}

We say $\gga$~is an \textit{operational isomorphism}, if it is a bijective homomorphism. In the next three lemmas, we prove that operational isomorphisms are exactly the (relational) isomorphisms (i.e., bijective embeddings) defined above. 

\begin{lemma}\label{lem:surj-embed-is-op-iso}
Every surjective embedding is an operational isomorphism.
\end{lemma}

\begin{myproof}
Pick any surjective embedding $\gga:U\to V$. We know $\gga$~is a relational isomorphism (Lemma~\ref{lem:surj-embed-is-rel-iso}), and thus bijective. So, we just need to show that it also respects union. For this, we let $X\subs U$ and prove that $\gga(\unicup X)=\bigcup_{u\in X}\gga(u)$. 

[$\sups$] Every $u\in X$ satisfies $u\subs \unicup X$ (clearly), and thus $\gga(u)\subs\gga(\unicup X)$ (since $\gga$~respects inclusion). Therefore the union of all $\gga(u)$ is also included in $\gga(\unicup X)$. 

[$\subs$] Since $V$~contains the $\gga(u)$ for all $u\in X$ and is a lattoid, it also contains their union $v\z:=\bigcup_{u\in X}\gga(u)$. Since $\gga$~is surjective, there exists $u\z\in U$ such that $\gga(u\z)=v\z$. Now, for every~$u\in X$, it is $\gga(u)\subs v\z$ (clearly); namely $\gga(u)\subs\gga(u\z)$; and thus $u\subs u\z$ (since $\gga$~respects non-inclusion). Therefore, the union $\unicup X$ of all such~$u\in X$ is also in~$u\z$, namely $\unicup X\subs u\z$. But then $\gga(\unicup X)\subs\gga(u\z)$ (since $\gga$~respects inclusion); namely $\gga(\unicup X)\subs v\z$, as desired.
\end{myproof}

\begin{lemma}\label{lem:inject-homom-is-embed}
Every injective homomorphism is an embedding. 
\end{lemma}

\begin{myproof}
Pick any injective homomorphism $\gga:U\to V$. Since $\gga$~respects inclusion (Lemma~\ref{lem:homoms-are-monotone}), we just need to show that it also respects non-inclusion. Let $u,u'\in U$ with $u\nsubs u'$. Then clearly $u\cup u'\nsubs u'$, and thus $u\cup u'\neq u'$; so, $\gga(u\cup u')\neq\gga(u')$ (since $\gga$~is injective); and thus $\gga(u)\cup\gga(u')\neq\gga(u')$ (since $\gga$~is a homomorphism); which is possible only if $\gga(u)\nsubs\gga(u')$, as desired. 
\end{myproof}

\begin{lemma}\label{lem:op-vs-rel-iso}
Operational and relational isomorphisms between lattoids coincide.
\end{lemma}

\begin{myproof}
Pick any $\gga:U\to V$. 
If $\gga$~is a relational isomorphism, then it is also an operational one (by Lemma~\ref{lem:surj-embed-is-op-iso}, as a surjective embedding). If $\gga$~is an operational isomorphism, then it is bijective (by definition); and an embedding (by Lemma~\ref{lem:inject-homom-is-embed}, as an injective homomorphism); hence, it is a relational isomorphism.
\end{myproof}

So, from now on we stop qualifying isomorphisms as relational or operational; and take it for granted that inclusion, non-inclusion, and union are all respected.

\subsection{Prime Volume under Surjective Homomorphisms}
\label{sec:prime-volume-under-surj-homom}

We now prove that surjective homomorphisms between lattoids never increase the prime volume (Lemma~\ref{lem:surj-homom-weakly-decreases-prime-vol}); and always decrease it, if they are non-injective and map to a regular lattoid (Lemma~\ref{lem:noninj-surj-homom-strictly-decreases-prime-vol}).

\begin{lemma}\label{lem:surj-homom-surjects-core}
If $\gga:U\to V$ is a surjective homomorphism, then every prime of~$V$ is the image of some prime of~$U$, and likewise for prime incomparabilities:
\begin{gather*}\label{equ:surj-homom-surjects-primes}
(\forall v\in\core V)
(\exists u\in\core U)
( \gga(u)=v ) 
\\\label{equ:surj-homom-surjects-prime-incomps}
\bigl( \forall \{v\kkk v'\}\in I(\core V) \bigr)
\bigl( \exists \{u\kkk u'\}\in I(\core U) \bigr)
\bigl( \{\gga(u), \gga(u')\}=\{v\kkk v'\} \bigr)\,.
\end{gather*}
\end{lemma}

\begin{myproof}
Pick any $v\in\core V$. Since $\gga$~is surjective, there is $u\in U$ with $\gga(u)=v$. Let~$X_u$ be any prime factorization of~$u$ (Lemma~\ref{lem:prime-factorization-exists}): $X_u\subs\core U \amp u=\unicup X_u$. We claim that some $u'\in X_u$ is the prime we are after, i.e., $\gga(u')=v$. Indeed: Let $Y:=\{ \gga(u') \mid u'\in X_u \}$ and use the fact that $\gga$~is a homomorphism, to get:
\begin{equation}
v=\gga(u)=\gga(\unicup X_u)=\!\!\bigcup_{u'\in X_u}\!\!\gga(u') = \unicup Y \,,
\end{equation}
namely $v=\unicup Y$ for the particular $Y\subs V$. Since $v$~is prime, it must be $v\in Y$, namely $v=\gga(u')$ for some $u'\in X_u$, as promised.

Now pick any prime incomparability $\{v\kkk v'\}$ of~$V$. By the previous part, there exist primes $u,u'$ in~$U$ with $\gga(u)=v$ and $\gga(u')=v'$. We claim $\{u\kkk u'\}$ is a prime incomparability of~$U$. Indeed: If not, then $u,u'$ are comparable. Without loss of generality, say $u\subs u'$. But then $\gga(u)\subs\gga(u')$ (as $\gga$~respects inclusion, by Lemma~\ref{lem:homoms-are-monotone}); namely $v\subs v'$; so $v,v'$ are comparable ---contradiction. 
\end{myproof}

\begin{lemma}\label{lem:surj-homom-weakly-decreases-prime-vol}
If $\gga:U\to V$ is a surjective homomorphism, then $U$~has at least as many primes and at least as many prime incomparabilities as~$V$:
\begin{equation}\label{equ:surj-homom-weakly-decreases-prime-vol:1}
|\core U|\;\geq\;|\core V|
\quad\text{and\/}\quad 
|I(\core U)|\;\geq\;|I(\core V)| 
\end{equation}
and thus also at least as large a prime volume: $\volume(\core U)\,\geq\,\volume(\core V)$.
\end{lemma}

\begin{myproofnoqed}
By Lemma~\ref{lem:surj-homom-surjects-core}, the restriction~$\hat\gga$ of~$\gga$ to~$\core U$ surjects it onto some $\hat V\sups\core V$; and the restriction~$\hat\ggi$ of the mapping $\{u\kkk u'\}\mapsto\{\gga(u),\gga(u')\}$ to $I(\core U)$ surjects it onto some $\hat I\sups I(\core V)$. Because of these surjections and inclusions, $|\core U|\geq|\hat V|\geq|\core V|$ and $|I(\core U)|\geq|\hat I|\geq|I(\core V)|$. Therefore,
$|\core U|\geq|\core V|$ and $|I(\core U)|\geq|I(\core V)|$; and
\begin{align}\label{equ:surj-homom-weakly-decreases-prime-vol:2}
\volume(\core U) 
&  =  |\core U|+|I(\core U)| \geq  \\
&\geq |\core V|+|I(\core V)| = \volume(\core V) \,. \tag*{\qed}
\end{align}
\end{myproofnoqed}

\begin{lemma}\label{lem:surj-homom-on-equal-size-cores}
Let $\gga:U\to V$ be a surjective homomorphism. If $U,V$ have equally many primes, then the restriction of~$\gga$ to~$\core U$ is a bijection to~$\core V$.\endnotemark
\end{lemma}
\myendnoteproof{Lemma}{\ref{lem:surj-homom-on-equal-size-cores}}{
As in the proof of Lemma~\ref{lem:surj-homom-weakly-decreases-prime-vol}, the restriction~$\hat\gga$ of~$\gga$ to~$\core U$ surjects it onto some $\hat V\sups\core V$ and thus $|\core U|\geq|\hat V|\geq|\core V|$. Since $|\core U|=|\core V|$, it must be that $\hat V=\core V$ and it must be that $\hat\gga$~is injective. Overall, $\gga$~is a bijection from $\core U$ onto~$\core V$.}

\begin{lemma}\label{lem:noninj-surj-homom-strictly-decreases-prime-vol}
Let $\gga:U\to V$ be a surjective homomorphism. If $\gga$~is non-injective and $V$~is regular, then\/ 
$\volume(\core U)>\volume(\core V)$.
\end{lemma}

\begin{myproof}
Let $\gga:U\to V$ and $V$ be as in the statement. By Lemma~\ref{lem:surj-homom-weakly-decreases-prime-vol}, we know \eqref{equ:surj-homom-weakly-decreases-prime-vol:1} holds; and it suffices to prove at least one of those inequalities strict (because then a calculation as in~\eqref{equ:surj-homom-weakly-decreases-prime-vol:2} shows that $\volume(\core U)>\volume(\core V)$). Equivalently, we prove that, if the first inequality of~\eqref{equ:surj-homom-weakly-decreases-prime-vol:1} is not strict, then the second one is. 

So, suppose $|\core U|=|\core V|$. To prove that $|I(\core U)|>|I(\core V)|$, we recall (from the proof of Lemma~\ref{lem:surj-homom-weakly-decreases-prime-vol}) the mapping $\{u\kkk u'\}\mapsto\{\gga(u),\gga(u')\}$ and its restriction~$\hat\ggi$ to~$I(\core U)$, which surjects it onto some $\hat I\sups I(\core V)$, causing $|I(\core U)|\geq|\hat I|\geq|I(\core V)|$. We prove that some prime incomparability $\{x\z,y\z\}$ is mapped by~$\hat\ggi$ to a two-set in~$\hat I\setminus I(\core V)$. This implies that $\hat I$~is strictly bigger than $I(\core V)$, and thus so is~$I(\core U)$.

To locate the desired $x\z,y\z$, we work as follows.

Since $\gga$ is non-injective, there exist $u,v\in U$ such that $u\neq v$ but $\gga(u)=\gga(v)$. We know $u\nsubs v \vee u\nsups v$ (as $u\neq v$), so let us assume (without loss of generality) that $u\nsubs v$. We also know $u,v$ have prime factorizations (Lemma~\ref{lem:prime-factorization-exists}), so let us fix them as $X,Y\subs\core U$ with $u=\unicup X$ and $v=\unicup Y$. Overall, we have:
\begin{equation}\label{equ:X,Y}
X,Y\subs\core U
\quad\text{and}\quad
\unicup X\nsubs\unicup Y
\quad\text{and}\quad
\gga(\unicup X)=\gga(\unicup Y) \,.
\end{equation}
To locate $x\z$, we fix any witness $i\z\in\unicup X\setminus\unicup Y$ of the non-inclusion in~\eqref{equ:X,Y}, and let~$x\z$ be any prime in~$X$ that contains it: $i\z\in x\z$. We note that, because of this witness, $x\z$ `escapes' all primes in~$Y$:
\begin{equation}\label{equ:x*-nsubs-Y}
\text{for all }y\in Y:\quad x\z\nsubs y \,.
\end{equation}
To locate~$y\z$, we observe that $\gga(x\z)$ is prime in~$V$ (by Lemma~\ref{lem:surj-homom-on-equal-size-cores}) and is thus represented (since $V$~is regular), by some~$r\z$; we see that this~$r\z$ is in
\begin{equation}
      \gga(x\z)
\subs \bigcup_{x\in X}\gga(x) 
 =    \gga(\unicup X) 
 =    \gga(\unicup Y)
 =    \bigcup_{y\in Y}\gga(y)
\end{equation}
(by the equality in~\eqref{equ:X,Y}); and let~$y\z$ be any prime in~$Y$ whose $\gga$-image contains it: $r\z\in\gga(y\z)$. We note that $y\z\neq x\z$, and in particular 
\begin{equation}\label{equ:x*-nsubs-y*}
x\z\nsubs y\z   
\end{equation}
(by~\eqref{equ:x*-nsubs-Y}); that $\gga(y\z)$~is also prime in~$V$ (again by Lemma~\ref{lem:surj-homom-on-equal-size-cores}); and that, since it contains the representative~$r\z$ of~$\gga(x\z)$, it also includes it entirely: 
\begin{equation}\label{equ:α(x*)-subs-α(y*)}
\gga(x\z)\subs\gga(y\z) \,.
\end{equation}
This further implies that the converse of this inclusion, $\gga(x\z)\sups\gga(y\z)$, is false (or else we would have $\gga(x\z)=\gga(y\z)$ for the distinct primes~$x\z,y\z$, contradicting Lemma~\ref{lem:surj-homom-on-equal-size-cores}). Hence, $\gga(x\z)\nsups\gga(y\z)$, which (by the monotonicity of~$\gga$) implies: 
\begin{equation}\label{equ:x*-nsups-y*}
x\z\nsups y\z \,.  
\end{equation}
At this point, we are done: our located primes $x\z,y\z$ are not comparable (by~\eqref{equ:x*-nsubs-y*} and~\eqref{equ:x*-nsups-y*}), but their (prime) images $\gga(x\z),\gga(y\z)$ are (by~\eqref{equ:α(x*)-subs-α(y*)}), and thus their two-set is not a prime incomparability in~$V$: $\{\gga(x\z),\gga(y\z)\}\not\in I(\core V)$.
\end{myproof}

\section{Connectivities}
\label{sec:connectivities}

Let $\bbB:=\{0\kkk1\}$ be the Boolean values. A \textit{connectivity} over~$h$ is any $h\times h$ \textit{Boolean matrix} $C=(c_{ij})_{1\leq i,j\leq h}\in\bbB^{h\times h}$. Equivalently, we think of~$C$ as the \textit{symbol} of~$\ggS_h$ where node~$i$ on the left connects to node~$j$ on the right iff $c_{ij}=1$ (Fig.\,\ref{fig:owl}); or as the \textit{mapping} $C:[h]\to\ppows h$ of every~$i\in[h]$ to the set $C(i):=\{j\in[h]\mid c_{ij}=1\}$.

In the latter case, we also consider the natural extension $\tilde C:\ppows h\to\ppows h$ which maps every set $u\subs[h]$ to $\tilde C(u):=\bigcup_{i\in u}C(i)$; and let $V(\tilde C):=\{\tilde C(u)\mid u\subs[h]\}$ be its range. Easily, $\tilde C$~is a \textit{homomorphism};%
\myendnote{Pick any collection of sets $X\subs\ppows h$. Then 
\begin{equation}\textstyle
\tilde C(\unicup X) 
\stackrel{(1)}{=} \bigcup_{i\in\unicup X} C(i)
\stackrel{(2)}{=} \bigcup_{u\in X} \bigl( \bigcup_{i\in u} C(i) \bigr)
\stackrel{(3)}{=} \bigcup_{u\in X} \tilde C(u) \,,
\end{equation}
where (1) and (3) use the definition of~$\tilde C$ out of~$C$; and (2)~simply reorganizes the union over all $i\in\unicup X$ to a union over all $i\in u$ over all~$u\in X$. Overall, we have $\tilde C(\unicup X)=\bigcup_{u\in X}\tilde C(u)$, as required for $\tilde C$ to be a homomorphism.}
and thus $V(\tilde C)$ is a \textit{lattoid} over~$h$ (Lemma~\ref{lem:range-of-homom-is-lattoid}). Importantly, $V(\tilde C)$ has at most~$h$ primes:

\begin{lemma}\label{lem:at-most-h-primes}
Let $C\in\bbB^{h\times h}$. For every prime $v\in V(\tilde C)$ there exists $i\in[h]$ such that $C(i)=v$. Therefore, $|\core{V(\tilde C)}|\leq h$.
\end{lemma}

\begin{myproof}
Let $v\in V(\tilde C)$ be prime. Since $v\in V(\tilde C)$, there is $u\subs[h]$ with $\tilde C(u)=v$. Let $X:=\{C(i)\mid i\in u\}$ collect the images of all elements of~$u$. Then $v=\tilde C(u)=\bigcup_{i\in u}C(i)=\unicup X$. But $v$~is prime, so it must be $v\in X$. So, there exists $i\in u$ such that $v=C(i)$. Overall, $C$~is a surjection from~$[h]$ onto a superset of the primes of~$V(\tilde C)$, causing $h=|[h]|\geq|\core{V(\tilde C)}|$.
\end{myproof}

The \textit{product} of two connectivities $A,B$ over~$h$ is the connectivity $C=AB$ over~$h$ obtained by \textit{Boolean multiplication} of matrices $A,B$, i.e., by the variant of matrix multiplication where $+$~and~$\cdot$ are replaced by $\vee$ and~$\wedge$, respectively:
\begin{equation}\label{equ:C=AB-as-matrix}
\text{for $i,j\in[h]$:}\quad 
c_{ij} := \textstyle\bigvee_{k=1}^h (a_{ik}\wedge b_{kj}) \,.
\end{equation}
Equivalently, we think of $C=AB$ as the \textit{concatenation} of the symbols~$A,B$ of~$\ggS_h$ (Fig.\,\ref{fig:owl}); or as the homomorphism obtained by standard \textit{function composition} of homomorphisms $A,B$, namely $C(u)=B(A(u))$, for all $u\subs[h]$.%
\footnote{\label{foot:overloading}For simplicity, from now on, we write~$C$ both for the mapping $C:[h]\to\ppows h$ and its extension $\tilde C:\ppows h\to\ppows h$. So, for $i\in[h],u\subs[h]$, notations $C(i),C(u)$ are both legal.}

\begin{lemma}\label{lem:α_A,B}
If $C=AB$ then the restriction of homomorphism $B:\ppows h\to\ppows h$ to $V(A)$ is a surjective homomorphism $\gga_{A,B}:V(A)\to V(C)$\textup; and $V(C)\subs V(B)$.
\end{lemma}

\begin{myproof}
In the composition of homomorphisms $A,B$ that produces~$C$, the second homomorphism~$B$ is applied only on the range of the first homomorphism~$A$ (i.e., on the lattoid~$V(A)$); and its respective values, which constitute the range of~$C$ (i.e., the lattoid~$V(C)$), are all in the range of~$B$ (i.e., in the lattoid~$V(B)$). Overall, the restriction~$\gga_{A,B}$ of homomorphism~$B$ to lattoid~$V(A)$ is a surjective homomorphism from~$V(A)$ onto~$V(C)$; and this last lattoid is in~$V(B)$.        
\end{myproof}

\subsection{Regularity}
\label{sec:regularity}

A connectivity~$C'$ is \textit{regular on} a connectivity~$C$ if there is connectivity~$J$ such that $C'=CJC$. Easily, this 
implies a surjective homomorphism inside~$V(C)$.
\begin{lemma}\label{lem:reg-implies-surj-homom}
If $C'$~is regular on~$C$ then there exists a surjective homomorphism $\gga:V(C)\to V(C')$\textup; and\/ $V(C')\subs V(C)$.
\end{lemma}

\begin{myproof}
Suppose $C'$~is regular on~$C$. Then $C'=CJC$, for some~$J$. Lemma~\ref{lem:α_A,B} for $C'=C(JC)$ says that $\gga:=\gga_{C,JC}$~is a surjective homomorphism from~$V(C)$ onto~$V(C')$. The same lemma for $C'=(CJ)C$ also says that $V(C')\subs V(C)$. 
\end{myproof}

Sometimes a connectivity~$C$ is regular on itself, i.e., $C=CJC$ for some~$J$. Then we just call it \textit{regular}. This is intimately related to regularity of lattoids.

\begin{lemma}\label{lem:reg-conn=reg-lattoid}
A connectivity $C$ is regular iff the lattoid $V(C)$~is regular.%
\footnote{Regular connectivities are also characterized by Zarecki\u\i's Theorem~\cite{za63,ya69}.}
\end{lemma}

\begin{myproof}

[$\Rightarrow$] Suppose $C=CJC$, for some~$J$. To show that~$V:=V(C)$ is regular, we fix an arbitrary prime~$v$ in~$V$ and show that it is represented.

By Lemma~\ref{lem:α_A,B}, the restriction~$\gga:=\gga_{C,JC}$ of homomorphism~$JC:\ppows h\to\ppows h$ to lattoid~$V(C)$ is a surjective homomorphism from~$V(C)$ to $V(CJC)=V(C)$, i.e., from~$V$ to~$V$. Hence (Lemma~\ref{lem:surj-homom-on-equal-size-cores}), its further restriction to~$\core V$ is a permutation of the primes of~$V$. In particular, $\tilde v:=\gga(v)$ is also a prime of~$V$. 

Now let $X:=\{\{i\}\mid i\in v\}$ be all singletons in~$v$, so that $v=\unicup X$; and $\tilde X:=\{JC(i)\mid i\in v\}=\{C(J(i))\mid i\in v\}$ be all their images in~$V(C)=V$. Then
\begin{equation}
\tilde v 
= \gga(v) 
= \gga(\unicup X)
= \gga(\bigcup_{i\in v} \{i\})
= JC(\bigcup_{i\in v} \{i\})
= \bigcup_{i\in v} JC(i)
= \unicup \tilde X \,.
\end{equation}
So, $\tilde v$ is the union of some sets~$\tilde X$ in~$V$. Since it is prime, we know $\tilde v\in\tilde X$, namely $\gga(v)=JC(i\z)$, from some $i\z\in v$. We claim $i\z$ is the representative we want.

For this, pick arbitrary prime~$u$ in~$V$ and suppose~$i\z\in u$. Then $\{i\z\}\subs u$, thus $JC(i\z)\subs JC(u)$ (since $JC$~is monotone, by Lemma~\ref{lem:homoms-are-monotone}), namely $\gga(v)\subs\gga(u)$. If $\gga$~respects non-inclusion, then this  implies $v\subs u$ and we are done. 

To see why $\gga$~respects non-inclusion, recall that it is a surjective homomorphism from~$V$ to~$V$. As a surjection between sets of equal size, it is injective, and thus a bijection. So, it is a bijective homomorphism, i.e., an operational isomorphism, and thus a relational one (Lemma~\ref{lem:op-vs-rel-iso}). Hence, it respects non-inclusion.

[$\Leftarrow$] Suppose every prime in $V:=V(C)$~has a representative. To show that $C$ is regular, we define a connectivity~$J$ and prove that $CJC=C$. 

We define~$J$ as the homomorphism $J=\tilde J:\ppows h\to\ppows h$ (cf.~Footnote~\ref{foot:overloading}) that extends to~$\ppows h$ the simpler mapping $J:[h]\to\ppows h$ of every $i\in[h]$ to the set
\begin{equation}\label{equ:J(i)}
J(i):=
\begin{cases}
\{ r \mid C(r)=v \} &\text{if $i$~represents some prime $v$ in~$V$,} \\
\emptyset           &\text{if $i$~represents no prime in~$V$.} \\
\end{cases}
\end{equation}
Recall that (i)~every $i$~can represent at most one prime 
(Endnote~\ref{end:each-rep-at-most-one-prime}); 
and (ii)~every prime is the image of some singleton (Lemma~\ref{lem:at-most-h-primes}); hence, if the upper case of~\eqref{equ:J(i)} applies, the prime~$v$ is unique and the set $J(i)$ is non-empty.

To prove $CJC=C$, it suffices that $JC$ is the identity~$I:V\to V$ (as then $CJC,C$ agree on all $u\in\ppows h$: $CJC(u)=JC(C(u))=I(C(u))=C(u)$). In turn, to prove that $JC=I$, it suffices that $JC$ is the identity on $\core V$ (Lemma~\ref{lem:values-on-core-determine-homomorphism}). 

So, let~$v$ be any prime. Consider its partition $v=v_0\cup v_1\cup v_2$ to its representatives, the representatives of other primes, and its non-representing elements: 
\begin{align}
v_0 &:= \{i\in v\mid i\text{ represents }v\} \\
v_1 &:= \{i\in v\mid i\text{ represents some prime }u\neq v\} \\
v_2 &:= \{i\in v\mid i\text{ represents no prime}\} \,.
\end{align}
Note that $v_0\neq\emptyset$ (as $V$~is regular); every $i\in v_1$ represents some~$u\subs v$ (since $v$~contains the representative~$i$ of~$u$); and $JC(v)=JC(v_0)\cup JC(v_1)\cup JC(v_2)$ (since $JC$~is a homomorphism). We claim that 
\begin{equation}\label{equ:J(v_i)}
JC(v_0)=v
\quad\text{and}\quad
JC(v_1)\subs v
\quad\text{and}\quad
JC(v_2)=\emptyset\,,
\end{equation}
and thus their union is indeed $JC(v)=v$, as required.

Indeed, every $i\in v_0$ represents~$v$, so $J(i)=\{ r \mid C(r)=v \}$; therefore
\begin{equation}\label{equ:JC(i)}
JC(i)
= C(J(i)) 
= C(\{ r \mid C(r)=v \}) 
= \hspace{-1ex}\bigcup_{C(r)=v}\hspace{-1.5ex} C(r) 
= \hspace{-1ex}\bigcup_{C(r)=v}\hspace{-1.5ex} v \;
= \; v \,,
\end{equation}
causing $JC(v_0) = \bigcup_{i\in v_0} JC(i) = \bigcup_{i\in v_0} v = v$ (also using that $v_0\neq\emptyset$). Similarly, every $i\in v_1$ represents some prime~$u\subs v$; so, $JC(i)\subs v$, and thus $JC(v_1)\subs v$; and every $i\in v_2$ represents no prime; so, $JC(i)=\emptyset$ and thus $JC(v_2)=\emptyset$.%
\myendnote{Indeed: For the arbitrary $i\in v_1$, we have:
\begin{equation}
JC(i)
= C(J(i)) 
= C(\{ r \mid C(r)=u \}) 
= \hspace{-1ex}\bigcup_{C(r)=u}\hspace{-1.5ex} C(r) 
= \hspace{-1ex}\bigcup_{C(r)=u}\hspace{-1.5ex} u \;
\subs \; v \,,
\end{equation}
which causes $JC(v_1) = \bigcup_{i\in v_1} JC(i) \subs \bigcup_{i\in v_1} v \subs v$ (even when $v_1=\emptyset$). Finally, the arbitrary $i\in v_2$, represents no prime, hence $J(i)=\emptyset$ and so
\begin{equation}
JC(i)
= C(J(i)) 
= C(\emptyset) 
= \emptyset\,, 
\end{equation}
causing $JC(v_2) = \bigcup_{i\in v_2} JC(i) = \bigcup_{i\in v_2} \emptyset = \emptyset$.}
\end{myproof}

Sometimes the relation between two connectivities is stronger than regularity: We say $C'$~is \textit{forward-regular on}~$C$ if there exists~$J$ such that $C'=C'JC=CJC$. Clearly, forward-regularity between connectivities implies regularity.

\begin{lemma}\label{lem:f-reg-implies-noninj-surj-homom}
If $C'$~is forward-regular on~$C$ and $C'\neq C$ then there exists a \emph{non-injective} surjective homomorphism $\gga:V(C)\to V(C')$.
\end{lemma}

\begin{myproofnoqed}
Suppose $C'$~is forward-regular on~$C$, so $C'=C'JC=CJC$ for some~$J$. As in the proof of Lemma~\ref{lem:reg-implies-surj-homom}, $C'=CJC$ implies that the restriction $\gga:=\gga_{C,JC}$ of homomorphism~$JC$ to the range~$V:=V(C)$ of~$C$ is a surjective homomorphism from~$V$ to~$V(C')$, and $V(C')\subs V$. We will prove that this $\gga$~is also non-injective.

Towards a contradiction, suppose $\gga$~is injective. Then it is a bijection from~$V$ to its subset~$V(C')$. But this can be only if $V(C')=V$. So, $\gga$~is a permutation of~$V$. Let~$k$ be its order, so that $\gga^k$~is the identity on~$V$. Then the homomorphism $C(JC)^k$ equals both~$C'$ and~$C$ (as shown below), hence $C'=C$ ---contradiction.

The equality $C(JC)^k=C'$ follows from the following stronger fact, for $t=k$.

\begin{claim}\label{clm:C'=C(JC)^t}
For all $t\geq1$: $C'=C(JC)^t$.
\end{claim}

\begin{mypproof}
By induction. If $t=1$, $C'=CJC$ by the choice of~$J$. If $t\geq2$, assume $C'=C(JC)^{t-1}$ and calculate: $C'=C'JC=\bigl(C(JC)^{t-1}\bigr)JC=C(JC)^t$.
\end{mypproof}

The equality $C(JC)^k=C$ holds because the two homomorphisms agree on all $u\in\ppows h$. Indeed, 
$
\bigl( C(JC)^k \bigr)(u) = 
(JC)^k\bigl( C(u) \bigr) = 
\gga^k\bigl( C(u) \bigr) = 
C(u) 
$,
where the second step uses the next claim, for $t=k$, and the fact that $C(u)\in V$; and the last step uses the fact that $\gga^k$ is the identity on~$V$ and indeed $C(u)\in V$.

\begin{claim}\label{clm:(JC)^t(v)=α(V)}
If~$v\in V$ then for all $t\geq1$: $(JC)^t(v) = \gga^t(v)$.
\end{claim}

\begin{mypproofnoqed}
By induction on~$t$. If $t=1$, then $JC(v) = \gga(v)$, since $\gga$~restricts~$JC$ to~$V$ and $v\in V$. If $t\geq2$, assume $(JC)^{t-1}(v)=\gga^{t-1}(v)$ and calculate:
\begin{equation}
(JC)^t(v) =
(JC)\bigl( (JC)^{t-1}(v) \bigr) \stackrel{(1)}{=}
(JC)\bigl( \gga^{t-1}(v) \bigr) \stackrel{(2)}{=}
\gga\bigl( \gga^{t-1}(v) \bigr) =
\gga^t(v) \,,
\end{equation}
where (1)~uses the inductive hypothesis; and (2)~is because $\gga$~restricts $JC$ to~$V$ and indeed $\gga^{t-1}(v)\in V$.\qqed$\Box$
\end{mypproofnoqed}
\end{myproofnoqed}

\section{Properties}
\label{sec:connectivity-properties}

The connectivity $C(z)$ of a string $z\in\ggS_h\s$ is the connectivity~$C\in\bbB^{h\times h}$ that records all connections between the two outer columns of~$z$, in that $c_{ij}=1$ iff $z$~has a one-way path from node~$i$ of the leftmost column to node~$j$ of the rightmost column (Fig.\,\ref{fig:owl}({\textsf{\scriptsize R}})). Easily, $z$~is live iff its connectivity is non-zero: $z\in\owl_h\iff C(z)\neq 0$; and concatenation of strings corresponds to Boolean multiplication of connectivities, namely $C(xy) = C(x)C(y)$. 

Given~$C\in\bbB^{h\times h}$, we let $P(C):=\{z\in\ggS_h\s\mid C(z)=C\}$ be all strings with connectivity~$C$. A property $P\subs\ggS_h\s$ is a \textit{connectivity property} if $P=P(C)$ for some~$C$. Easily, connectivity properties are non-empty: $P(C)\neq\emptyset$, for all~$C$. Moreover, when properties are connectivity properties, some qualities and relations of properties correspond to qualities and relations of connectivities.

\begin{lemma}\label{lem:conn-props-to-conns}
Let $C,C'\in\bbB^{h\times h}$ be any two connectivities. Let $P:=P(C)$ and $P':=P(C')$ be the corresponding connectivity properties. Then\textup:\endnotemark
\begin{itemize}
\item[\textup{(a)}] $P$~is smooth iff\/ $C$~is regular. 
\item[\textup{(b)}] $P$~has suffix of choice into~$P'$ iff\/ $C'$~is forward-regular on~$C$.
\item[\textup{(c)}] $P, P'$ are separated by~$\owl_h$ iff\/ $C\neq C'$.
\end{itemize}
\end{lemma}
\myendnoteproof{Lemma}{\ref{lem:conn-props-to-conns}}{%
(a)~[$\Rightarrow$]~Suppose $P$~is smooth. Let $x,z\in P$, namely $C(x)=C(z)=C$. Using the smoothness of~$P$, pick any~$y$ such that $xyz\in P$, namely $C(xyz)=C$, and let $J:=C(y)$. Then $C=C(xyz)=C(x)C(y)C(z)=CJC$. Overall, we found~$J$ such that $C=CJC$. Hence, $C$~is regular.

[$\Leftarrow$]~Suppose $C$~is regular. Let $x,z\in P$, namely $C(x)=C(z)=C$. Since $C$~is regular, there is~$J$ such that $C=CJC$. We know $P(J)\neq\emptyset$, so pick any $y\in P(J)$. Then $C(xyz)=C(x)C(y)C(z)=CJC=C$, hence $xyz\in P$. Overall, we found $y$ such that $xyz\in P$. Hence, $P$~is smooth.

(b)~[$\Rightarrow$]~Pick any $v,x\in P$, $x'\in P'$. Then $C(v)=C(x)=C$ and $C(x')=C'$. Since $P$~has suffix of choice into~$P'$, there exists $u$ such that $xuv,x'uv\in P'$, namely $C(xuv)=C(x'uv)=C'$. Let $J:=C(u)$. Then  
\begin{gather}
C'=C(xuv)=C(x)C(u)C(v)=CJC    \,,\\
C'=C(x'uv)=C(x')C(u)C(v)=C'JC \,.   
\end{gather}
Overall, we found $J$ such that $C'=C'JC=CJC$, as required.

[$\Leftarrow$]~Suppose $C'$~is forward regular on~$C$. Then $C'=C'JC=CJC$ for some~$J$. Let $u\in P(J)\neq\emptyset$ be any string with that connectivity. Now pick any $v,x\in P$ and $x'\in P'$, namely $C(v)=C(x)=C$ and $C(x')=C'$. Then
\begin{gather}
C(xuv)=C(x)C(u)C(v)=CJC=C'    \,,\\
C(x'uv)=C(x')C(u)C(v)=C'JC=C' \,.   
\end{gather}
Overall, for any $v\in P$, we found $u$ with $xuv\in P$ for all $x\in P\cup P'$, as required.

(c)~[$\Rightarrow$]~For the contrapositive, suppose $C=C'$. Pick any $x\in P$ and $x'\in P'$. Then $C(x)=C=C'=C(x')$. Hence, for any context $u,v$, it is $C(uxv)=C(u)C(x)C(v)=C(u)C(x')C(v)=C(ux'v)$, namely $uxv$ and $ux'v$ have the same connectivity. Hence they are both dead (i.e., of connectivity zero) or both live (i.e., of connectivity non-zero). Hence, $\owl_h$ does not separate $P,P'$.

[$\Leftarrow$]~Suppose $C\neq C'$. Then there exist $1\leq i,j\leq h$ such that $c_{ij}\neq c'_{ij}$. Without loss of generality, assume $c_{ij}=1$ and $c'_{ij}=0$. Now pick any $x\in P$ and $x'\in P'$. Then $C(x)=C$ and $C(x')=C'$, which means that $x$~has a path from node~$i$ on the leftmost column to node~$j$ on the rightmost column, but $x'$~does not. Hence, if $u$~is the single symbol $\{(i,i)\}$ and $v$ the single symbol $\{(j,j)\}$,  then the context~$u,v$ distinguishes between $x$ and $x'$: $uxv$~is live and $ux'v$~is dead. Overall, $\owl_h$~separates $P,P'$.}

We now use our theory of lattoids and connectivities to prove our main result (Th.~\ref{thm:main}) \emph{in the special case where we use only connectivity properties} (as was the case in~\cite{adka26sofsem}), namely Th.~\ref{thm:main-conn} below. First, a final lemma: two connectivity properties that satisfy the conditions of \cite[Lemma~10]{adka26sofsem} decrease the prime volume.

\begin{lemma}\label{lem:main}
Let $P,P'\subs\ggS_h\s$ be two \emph{connectivity properties}. Suppose $P,P'$~are separated by~$\owl_h$\textup; $P$~has suffix of choice into~$P'$\textup; and $P'$~is smooth. Then the corresponding connectivities $C,C'$ satisfy\textup: $\volume(\core{V(C)})>\volume(\core{V(C')})$. 
\end{lemma}

\begin{myproof}
We know $C'$~is forward-regular on~$C$ (by Lemma~\ref{lem:conn-props-to-conns}b, since $P$~has suffix of choice into~$P'$) and $C\neq C'$ (by Lemma~\ref{lem:conn-props-to-conns}c, as $P,P'$~are separated by~$\owl_h$). So (Lemma~\ref{lem:f-reg-implies-noninj-surj-homom}), a non-injective surjective homomorphism $\gga:V(C)\to V(C')$ exists. Since $V(C')$~is also regular (by Lemmas~\ref{lem:reg-conn=reg-lattoid} \&~\ref{lem:conn-props-to-conns}a, as $P'$~is smooth), the prime volume decreases (Lemma~\ref{lem:noninj-surj-homom-strictly-decreases-prime-vol}):  $\volume(\core{V(C)})>\volume(\core{V(C')})$.
\end{myproof}

\begin{theorem}\label{thm:main-conn}
Let $P_0,P_1,\dots,P_m\subs\ggS_h\s$ be \emph{connectivity properties}. Suppose that, for~$1\leq i\leq m$\textup: 
$P_{i-1},P_i$~are separated by~$\owl_h$\textup; $P_{i-1}$~has suffix of choice into~$P_i$\textup; and $P_i$~is smooth. Then $m\leq\binom{h+1}{2}$.
\end{theorem}

\begin{myproof}
For $i=0,1,\dots,m$, let $C_i$~be the connectivity that underlies property~$P_i$, namely $P_i=P(C_i)$; and~$s_i,t_i$ be respectively the number of primes and the prime volume of the lattoid~$V(C_i)$, namely 
\begin{equation}
s_i:=|\core{V(C_i)}|
\qquad\text{and}\qquad
t_i:=\volume(\core{V(C_i)}) \,.  
\end{equation}
Then $s_i\leq h$ (Lemma~\ref{lem:at-most-h-primes}) and $0\leq t_i \leq\binom{s_i+1}{2}$ 
(Endnote~\ref{end:bounds-for-prime-volume}), 
so $0\leq t_i\leq \binom{h+1}{2}$; but also $t_{i-1}>t_i$ (by Lemma~\ref{lem:main}, applied to $P_{i-1},P_i$). So, the sequence $t_0,t_1,\dots,t_m$ contains $1+m$~distinct numbers between~$0$ and $\binom{h+1}{2}$. Hence, $m\leq\binom{h+1}{2}$.
\end{myproof}

\section{Conclusion}
\label{sec:conclusion}

In~\cite{adka26sofsem}, we proved a lower bound of $\frac12N$ for any \tdfa\ for~$\owl_h$, where $N=\binom{h+1}{2}$. At the heart of our proof lied a sequence of \textit{connectivity properties} $P_0,P_1,\dots,P_N$ meeting certain specifications. It was clear that a longer such sequence would imply a greater lower bound. So, we asked: \textit{Could a better design produce a longer sequence of connectivity properties, and thus a greater lower bound?} Here, we answered negatively, i.e., that the sequence designed in~\cite{adka26sofsem} was longest (Th.~\ref{thm:main-conn}). 

Importantly, our proof would have also worked with \textit{arbitrary properties}; we used connectivity ones only because our design took us there. So, one may wonder: \textit{What if we allow arbitrary properties? Is there a longer sequence then?} The answer is again negative (Th.~\ref{thm:main}), but requires a  longer presentation, which we are reserving for the full version of this report.  

This work used no artificial intelligence.


\newif\ifshortbibvenues

\newcommand{\newconfsl}[3]{\newcommand{#1}{Proceedings of \ifshortbibvenues #2\else the #3\fi}}
\newcommand{\newconfl }[2]{\newcommand{#1}{Proceedings of the #2}}

\newconfsl{\confCIAA  }{CIAA}  {{C}onference on {I}mplementation and {A}pplication of {A}utomata}
\newconfsl{\confCiE   }{CiE}   {{C}onference of {C}omputability in {E}urope}
\newconfsl{\confCCC   }{CCC}   {{C}onference on {C}omputational {C}omplexity}
\newconfsl{\confCSR   }{CSR}   {{I}nternational {C}omputer {S}cience {S}ymposium in {R}ussia}
\newconfsl{\confDCFS  }{DCFS}  {{W}orkshop on {D}escriptional {C}omplexity of {F}ormal {S}ystems}
\newconfsl{\confDLT   }{DLT}   {{I}nternational {C}onference on {D}evelopments in {L}anguage {T}heory}
\newconfsl{\confFCT   }{FCT}   {{I}nternational {S}ymposium on {F}undamentals of {C}omputation {T}heory}
\newconfsl{\confFOCS  }{FOCS}  {{S}ymposium on the {F}oundations of {C}omputer {S}cience}
\newconfsl{\confICALP }{ICALP} {{I}nternational {C}olloquium on {A}utomata, {L}anguages, and {P}rogramming}
\newconfsl{\confICIP  }{ICIP}  {{I}nternational {C}onference on {I}mage {P}rocessing}
\newconfsl{\confISTCS }{ISTCS} {{I}srael {S}ymposium on the {T}heory of {C}omputing {S}ystems}
\newconfsl{\confLATA  }{LATA}  {{L}anguage and {A}utomata {T}heory and {A}pplications}
\newconfsl{\confMBB   }{MBB}   {{G}raduate {S}tudent {C}onference on {T}he {N}ature of {T}hought}
\newconfsl{\confMFCS  }{MFCS}  {{I}nternational {S}ymposium on {M}athematical {F}oundations of {C}omputer {S}cience}
\newconfsl{\confNCMA  }{NCMA}  {{I}nternational {W}orkshop on {N}on-{C}lassical {M}odels for {A}utomata and {A}pplications}
\newconfsl{\confSAGA  }{SAGA}  {{I}nternational {S}ymposium on {S}tochastic {A}lgorithms: {F}oundations and {A}pplications}
\newconfsl{\confSOAM  }{SOAM}  {{S}ymposium on {A}pplied {M}athematics}
\newconfsl{\confSODA  }{SODA}  {{S}ymposium on {D}iscrete {A}lgorithms}
\newconfsl{\confSOFSEM}{SOFSEM}{{I}nternational {C}onference on {C}urrent {T}rends in {T}heory and {P}ractice of {I}nformatics}
\newconfsl{\confSTACS }{STACS} {{S}ymposium on {T}heoretical {A}spects of {C}omputer {S}cience}
\newconfsl{\confSTOC  }{STOC}  {{S}ymposium on the {T}heory of {C}omputing}
\newconfl {\confSWAT  }{{S}ymposium on {S}witching and {A}utomata {T}heory}
\newconfl {\confSWCT  }{{S}ymposium on {S}witching {C}ircuit {T}heory and {L}ogical {D}esign}
\newconfl {\confIMYCS }{{I}nternational {M}eeting of {Y}oung {C}omputer {S}cientists}

\newcommand{\bibvenueskey}{\ifbibvenuesshort{\footnotesize 
CIAA   is the \textit{Conference on Implementation and Application of Automata}.
CiE    is the \textit{Conference of Computability in Europe}
CCC    is the \textit{Conference on Computational Complexity}.
CSR    is the \textit{International Computer Science Symposium in Russia}.
DCFS   is the \textit{Workshop on Descriptional Complexity of Formal Systems}.
DLT    is the \textit{International Conference on Developments in Language Theory}.
FCT    is the \textit{International Symposium on Fundamentals of Computation Theory}.
FOCS   is the \textit{Symposium on the Foundations of Computer Science}.
ICALP  is the \textit{International Colloquium on Automata, Languages, and Programming}.
ICIP   is the \textit{International Conference on Image Processing}.
ISTCS  is the \textit{Israel Symposium on the Theory of Computing Systems}.
LATA   is the \textit{Conference on Language and Automata Theory and Applications}
MBB    is the \textit{Graduate Student Conference on The Nature of Thought}.
MFCS   is the \textit{International Symposium on Mathematical Foundations of Computer Science}.
SAGA   is the \textit{International Symposium on Stochastic Algorithms: Foundations and Applications}.
SOAM   is the \textit{Symposium on Applied Mathematics}.
SODA   is the \textit{Symposium on Discrete Algorithms}.
SOFSEM is the \textit{International Conference on Current Trends in Theory and Practice of Informatics}
STACS  is the \textit{Symposium on Theoretical Aspects of Computer Science}.
STOC   is the \textit{Symposium on the Theory of Computing}.
}\else\fi}
\shortbibvenuestrue 
\bibliographystyle{elsarticle-num}
\bibliography{biblio/string,biblio/collections,biblio/personal,biblio/computation}

\notesnameforsubmission
\theendnotes

\end{document}